\documentclass[
               twocolumn,
               noshowpacs,          
               nopreprintnumbers,     
               aps,                 
               prd,                 
               a4paper,             
               superscriptaddress,  
               nofootinbib,         
               tightenlines,        
               floats,floatfix      
               ]{revtex4}

\usepackage{amsmath, amsfonts, amsthm, amssymb, graphicx}
\usepackage{epstopdf}
 \usepackage{slashed}
 \usepackage{graphicx,epsfig}
\usepackage{amsmath}
\usepackage {amssymb}
\usepackage{caption}
\usepackage{subcaption}
\usepackage[utf8]{inputenc}
\usepackage{hyperref}
\usepackage[francais]{babel}
\usepackage[T1]{fontenc}
\usepackage{float}
\usepackage{xcolor}

\newcommand{\be}{\begin{eqnarray}}
\newcommand{\ee}{\end{eqnarray}}
\newcommand{\bea}{\begin{eqnarray}}
\newcommand{\eea}{\end{eqnarray}}

\begin{document}

\title{Triple Points of Gravitational AdS Solitons and Black Holes}

\author{Constanza Quijada}
\affiliation{Departamento de Física, Universidad de Concepción, Casilla, 160-C, Concepción, Chile}

\author{Andrés Anabalón}
\affiliation{Departamento de Física, Universidad de Concepción, Casilla, 160-C, Concepción, Chile}

\affiliation{Instituto de Física Teórica, UNESP-Universidade Estadual Paulista, R. Dr. Bento T. Ferraz 271, Bl.
II, Sao Paulo 01140-070, SP, Brazil.
}
\author{Robert B. Mann}
\affiliation{Department of Physics and Astronomy, University of Waterloo, Waterloo, Ontario, N2L 3G1, Canada}

\author{Julio Oliva}
\affiliation{Departamento de Física, Universidad de Concepción, Casilla, 160-C, Concepción, Chile}




\begin{abstract}
We present a triple point of a new kind for General Relativity, at which two gravitational solitons can coexist with a planar black hole in anti de Sitter space. Working in the context of
  non-linear electrodynamics, we obtain simple, sensible spacetimes for which the thermodynamics can be studied in an analytic manner. The spacetimes are charged under the non-linear electrodynamics leading to an electric charge for black holes and a magnetic flux for solitons. In the grand-canonical ensemble, we show that the phase space of the theory is very rich, containing re-entrant phase transitions, as well as triple points, for small values of the coupling controlling the non-linearity of the electrodynamics Lagrangian.
\end{abstract}

\maketitle

\section{Introduction}

Black holes in Anti-de Sitter (AdS) spacetime have thermal properties that may be structurally different than those of their asymptotically flat counterparts. Even in vacuum, the former may have positive specific heat, leading to locally stable thermal objects, suitable for exploring phase transitions. The Hawking-Page (HP) transition emerges in this manner, since the difference in free energy between large black holes and thermal AdS changes its sign at a given critical temperature $T_c$, above which the black hole dominates, while for $T<T_c$ thermal AdS is the preferred phase \cite{Hawking:1982dh}. This phase transition is particularly relevant in holography \cite{Maldacena:1997re}, since it leads to a confinement/deconfinement transition in the dual field theory \cite{Witten:1998zw}. When gauge fields are introduced, it is natural to consider charged solutions. In this context, in \cite{Chamblin:1999tk} the authors explored  the structure of the phase space in both the canonical and grand-canonical ensembles that accommodate charged, spherically symmetric black holes, and showed that the effect of the electric potential on the phase space structure is crucial. For large electric potential, there is a single black hole that always dominates the canonical ensemble. 
If thermodynamic pressure is taken into account, then black holes of fixed charge can undergo Van der Waals phase transitions  \cite{Kubiznak:2012wp}.

Furthermore, General Relativity with a negative cosmological constant naturally contains black holes with non-spherical horizons \cite{Mann:1997iz} that can be planar or hyperbolic, leading to non-trivial topologies that propagate up to the conformal boundary. Consequently these black holes are asymptotically locally AdS. In holography, planar black holes are of special relevance since they are dual to Conformal Field Theories (CFTs) at finite temperature formulated in the flat conformal boundary $\mathbb{R}_t \times \mathbb{R}^{d-1}$.
In \cite{Horowitz:1998ha}, the ground state of the space of planar black holes  was identified as a gravitational soliton, dubbed the AdS soliton. 
The latter can be obtained from a double analytic continuation of the planar black hole, and it therefore defines a new configuration that competes as a possible dominant thermal phase. Phase transitions between a planar black hole and the AdS soliton at finite temperature were originally explored in \cite{Surya:2001vj}, a study that was extended for charged black holes in \cite{Banerjee:2007by}, still within the context of the neutral soliton, which is consistent in the grand-canonical ensemble.

Recently  a new charged AdS-soliton configuration was discovered \cite{Anabalon:2021tua}. This solution of the Einstein-Maxwell system with a negative cosmological constant can be obtained from a double analytic continuation of the planar, electrically charged AdS black hole, which also requires  an analytic continuation of the electric charge, leading indeed to a smooth, charged soliton. Even more, when this model is embedded in gauge supergravity, both in dimensions four and five, the soliton turns out to be supersymmetric\footnote{These studies were extended to $\mathcal{N}=4$, $SU(2)\times SU(2)$ supergravity \cite{Freedman:1978ra}, in \cite{Canfora:2021nca}.}
for a particular value of the charge \cite{Anabalon:2022aig}.  The magnetic charge of the soliton leads to a magnetic flux, which enters as a control variable in a new ensemble, extending the results of \cite{Banerjee:2007by}, where only neutral solitons were considered. It turns out that comparing the corresponding free energy of different configurations at equal temperature, electric potential, magnetic flux and period of a spatial cycle, leads to an interesting structure of the phase space, since for large electric potential the black hole phase dominates \cite{Anabalon:2022ksf}.

It is well-known that for large electric fields, Maxwell's equations receive non-linear corrections coming from quantum effects induced by fermion loops, as well as from the low energy limit of String Theory. Furthermore, non-linear electrodynamics (NLED) naturally emerges inside materials with non-linear constitutive relations.  NLED was also employed to construct the first regular black holes of Einstein's theory coupled to a field theory \cite{Ayon-Beato:1998hmi}. With this motivation, here we commence an exploration of the effects of NLED on the thermal stability of charged solitons.

There are different modifications of Maxwell theory, which are well-motivated, depending on the particular setup one is considering \cite{Sorokin:2021tge}. 
We consider the most general form of
NLED introduced by Gao  \cite{Gao:2021kvr}
that leads to a simple correction of the gravitational and gauge potentials, allowing  analytic control in most of the study. Recently, the phase space of a family of NLED \cite{Gao:2021kvr}, 
was shown to posses multiple points at which four and even five different black hole branches with the same free energy \cite{Tavakoli:2022kmo} can coexist. 
Remarkably, we find  that in this simple setup a new sort of triple point can emerge, at which two solitons and one black hole branch can coexist.


\section{Black holes and Solitons in NLED}

We consider the framework defined by the theory  \cite{Gao:2021kvr} 
\begin{equation}
    S = \frac{1}{2\kappa} \int_\mathcal{M} d^4 x \ \sqrt{-g} \left( R - 2 \Lambda - L_{\mathrm{EM}} \right) \ , \label{action}
\end{equation}
where the electromagnetic Lagrangian is given by \cite{Gao:2021kvr}
\begin{equation}
    L_{\mathrm{EM}} = \sum_{i=1}^{\infty} \alpha_i \left( F^2 \right)^i \ , 
\end{equation} 
with $F^2=F_{\mu \nu} F^{\mu \nu}$ and $F_{\mu \nu}=\partial_\mu A_\nu - \partial_\nu A_\mu$. The $\alpha_i$ are dimensionful coupling constants ($[\alpha_i] = \text{Length}^{2(i-1)}$), and $A_\mu$ is the $U(1)$ Maxwell field. 

By varying the action with respect to the metric and the gauge fields, the following equations are obtained 
    \begin{eqnarray}
    R_{\mu \nu} - \frac{1}{2} g_{\mu \nu} R + \Lambda g_{\mu \nu} &=& T_{\mu \nu} \ , \label{eqg}\\
     \nabla_\mu \left( \frac{\partial L_\mathrm{EM}}{\partial F^2} F^{\mu \nu} \right) &=& 0 \label{eqA} \ , 
\end{eqnarray}
with 
\begin{equation}
    T_{\mu \nu} = -\frac{1}{2} g_{\mu \nu} L_\mathrm{EM}+2\frac{\partial L_\mathrm{EM}}{\partial F^2} F_{\mu \alpha} F_{\nu}^{\ \alpha} \ .
\end{equation}

Charged, static black hole spacetimes with planar horizons belong to the following family of metrics and gauge potentials
\begin{eqnarray}\label{metricandgaugebh}
    ds_b^2 &=& - U(r) dt_b^2 + \frac{dr^2}{U(r)} + \frac{r^2}{\ell^2} (d\varphi_b^2 + dz^2) \ , \\
    A_b &=& A_t (r) dt_b  + A_{\varphi} d\varphi_{b}\ ,
\end{eqnarray}
where we have chosen to identify $\varphi_b$ via $0\leq\varphi_b\leq \eta_b$, while $0<z<L$ and $A_{\varphi}$ is an arbitrary constant. 

Using this ansatz, the field equations \eqref{eqg} and \eqref{eqA} yield the relations 
\begin{eqnarray}
r U^{\prime \prime} + 2 U^\prime - \frac{6r}{\ell^2} + r \sum_{i=1}^{\infty} \alpha_i (-2)^i (A_t^\prime)^{2i} &=& 0 \ ,  \label{rel1} \\
\frac{r^2}{2} \sum_{i=1}^{\infty} i \alpha_i (-2)^i (A_t^\prime)^{2i-1} - Q &=& 0 \ , \label{rel2} 
\end{eqnarray} 
where the prime denotes derivative with respect to the radial coordinate $r$ and $Q$ is an integration constant corresponding to the electric charge of the black hole. 

To solve this system of equations we employ the expansions
\begin{equation}
    A_t (r) = \sum_{i=1}^{\infty} b_i r^{-i} \quad \mathrm{and} \quad U(r) = \frac{r^2}{\ell^2} + \sum_{i=1}^{\infty} c_i r^{-i} \ , \label{completos}
\end{equation}
where $b_i$ and $c_i$ are constants. Inserting these expressions in \eqref{rel1} and \eqref{rel2}, and setting $\alpha_1 = 1$ (thereby  recovering Maxwell theory when $\alpha_{i\geq2}=0$) yields 
\begin{eqnarray}
    b_1 &=& Q \ ,\\
    b_5 &=& \frac{4}{5} Q^3 \alpha_2 \ , \\ 
    b_9 &=& \frac{4}{3} Q^5 \left( 4 \alpha_2^2 - \alpha_3 \right) \ ,\\ 
    b_{13} &=& \frac{32}{13}Q^7 \left( 24 \alpha_2^3 - 12\alpha_3 \alpha_2 + \alpha_4 \right) \ , \\ 
    \dots &=& \dots \ ,
\end{eqnarray}
and
\begin{equation}
    c_1=-2M \quad c_i = 
    \frac{4Q}{i+2} b_{i-1} , \quad \mathrm{for} \quad i>1  \ , 
\end{equation}
where $M$ corresponds to the mass of the black hole. All the non-listed $b_i$ vanish.

The key point identified in \cite{Gao:2021kvr} is that 
there are infinitely many 
choices of  couplings $\alpha_i$ that yield  a finite number of non-vanishing $b_i$, and consequently $c_i$, and vice-versa. 
In such cases the simple structure of the gauge potential and metric functions allows  exploration of the phase space in an analytic manner.

From \eqref{action}, it is possible to obtain black holes solutions with 2, 3 or more event horizons, setting the $b_i$ constants to zero for sufficiently large $i$, consequently leading to relations between the $\alpha_i$. In the case of the 2-horizon black hole, the values of $b_i$ are set to zero for $i\geq 5$, implying  $\alpha_i=0$ for $i \geq 2$;  thus we recover  Einstein-Maxwell theory with cosmological constant, and the black hole solution reduces to the Reissner-Norstrom-AdS geometry.

The general form for a black hole solution is 
\begin{equation}
    A_t = \sum_{i=1}^{\infty} 
    \left(\frac{Q \tilde{b}_i}{r^{i}} -\frac{ Q \tilde{b}_i}{r_+^{i}}
    \right)
    \quad  U = \frac{r^2}{\ell^2}
 -\frac{2M}{r}   + \sum_{i=2}^{\infty}
 \frac{4Q^2 \tilde{b}_{i-1}}{(i+2) r^i} \label{planarbh}
\end{equation}
assuming $U(r)$ has a root $r=r_+$, where
${b}_{i}= Q\tilde{b}_{i}$.  The Euclidean on-shell action yields a particularly simple expression for the free energy (see section IV), in spite of the presence of an infinite series, yielding
\begin{align}
M &= \frac{r_+^3}{2\ell^2} + \sum_{i=2}^{\infty}
 \frac{2Q^2\;  \tilde{b}_{i-1}}{(i+2) r_+^{i-1}} \label{genM}\\
 T &= \frac{3r_+}{4 \pi \ell^2} - \sum_{i=2}^{\infty}
 \frac{Q^2\; \tilde{b}_{i-1}(i-1)}{ \pi (i+2) r_+^{i+1}} \label{genT}\\
\phi_b &=  \sum_{i=1}^{\infty} 
    \frac{Q\;  \tilde{b}_i}{r_+^{i}} \label{genphi}
\end{align}
for the mass, temperature, and electric potential of the black hole.


To obtain the general soliton solution, we 
apply a double Wick rotation to the planar black hole metric \eqref{metricandgaugebh} ($t_b \rightarrow i \varphi_s$ and $\varphi_b \rightarrow i t_s$) and  an analytic continuation of the black hole charge ($Q \rightarrow -iq$). This gives 
\begin{eqnarray}
    ds^2_s &=& - \frac{r^2}{\ell^2} dt_s^2 + \frac{dr^2}{U(r)} + U(r) d\varphi_s^2 + \frac{r^2}{\ell^2} dz^2 \\
    A_s &=& A_t dt_s + A_{\varphi}(r) d\varphi_s
\end{eqnarray}
where $0<z<L$,  $0\leq\varphi_s\leq \eta_s$, identified, and
\begin{equation}
    A_\varphi = \sum_{i=1}^{\infty} 
    \left(\frac{q \hat{b}_i}{r^{i}} -\frac{ q \hat{b}_i}{r_0^{i}}
    \right)
    \quad  U = \frac{r^2}{\ell^2}
 -\frac{2\mu}{r}   - \sum_{i=2}^{\infty}
 \frac{4q^2 \hat{b}_{i-1}}{(i+2) r^i} \label{solbh}
\end{equation}
 where
$\hat{b}_{i}(q)= \tilde{b}_{i}(iQ)$, and
\begin{align}
\mu &= \frac{r_0^3}{2\ell^2} - \sum_{i=2}^{\infty}
 \frac{2q^2\;  \hat{b}_{i-1}}{(i+2) r_0^{i-1}} \label{gensolM}\\
\Phi_b &=  \sum_{i=1}^{\infty} 
    \frac{Q\;  \hat{b}_i}{r_0^{i}} \eta_s\label{gensolphi}
\end{align}
are the respective mass  and magnetic flux of the soliton.  The period $\eta_s$ of the $\varphi_s$ coordinate of the soliton is fixed in terms of $r_0$ and $q$ as 
\begin{equation}
\eta_s  =  \left[ \frac{3r_0}{4 \pi \ell^2} + \sum_{i=2}^{\infty}
 \frac{q^2\; \hat{b}_{i-1}(i-1)}{ \pi (i+2) r_0^{i+1}}\right]^{-1}
\end{equation}
in order to avoid a potential conical singularity that may appear at the origin of the soliton, located at $r=r_0$.

\section{Free energy and phase transitions}

The on-shell Euclidean action receives three contributions
\begin{equation}\label{IEuc}
    I_\textrm{Euc} = -I_{\mathrm{bulk}} -I_{\mathrm{GH}} + I_{\mathrm{ct}}
\end{equation}
where 
\begin{eqnarray}
    I_{\mathrm{bulk}} &=& \frac{1}{2\kappa} \int_\mathcal{M} d^4 x \sqrt{g} \left( R - 2 \Lambda - \sum_{i=1}^{\infty} \alpha_i \left( F^2 \right)^i \right) \ , \\
    I_{\mathrm{GH}} &=& \frac{1}{\kappa} \int_{\partial\mathcal{M}} d^3 x \sqrt{h} \ K \ , \\
    I_{\mathrm{ct}} &=& \frac{1}{\kappa} \int_{\partial\mathcal{M}} d^3 x \sqrt{h} \left( \frac{2}{\ell} - \frac{\ell}{2} \mathcal{R}(h) \right) \ .
\end{eqnarray}
No further boundary term is required to compare the free energy at fixed temperature, electric potential, spacelike period and magnetic flux.
From \eqref{IEuc} we obtain 
\begin{eqnarray}
    G_b &=& - \frac{L \eta_b}{\kappa \ell^2} M \\
    G_s &=& - \frac{L \eta_s}{\kappa \ell^2} \mu  
\end{eqnarray}
for the free energy of the black hole and the soliton  
respectively.  
\begin{figure}[H]\label{fig:alphaverysmall}
     \centering
     \begin{subfigure}[b]{0.3\textwidth}
         \centering
         \includegraphics[width=\textwidth]{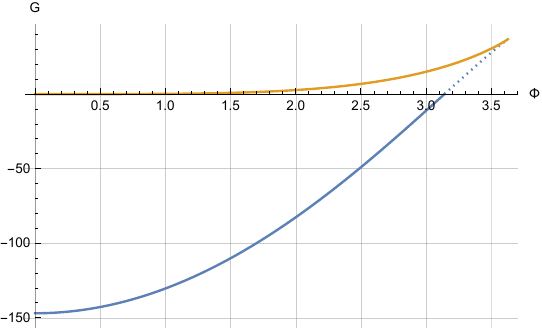}
         \caption{$\alpha_2=0.0$}
     \end{subfigure}
     \qquad
     \begin{subfigure}[b]{0.3\textwidth}
         \centering
         \includegraphics[width=\textwidth]{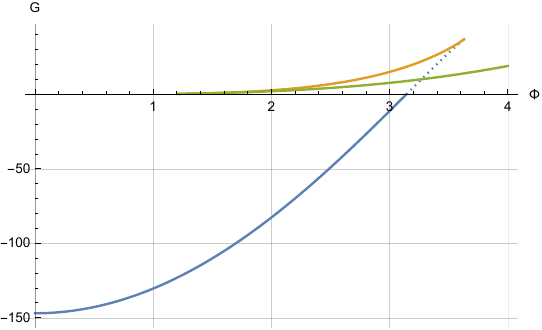}
         \caption{$\alpha_2=0.001$}
     \end{subfigure} \\
      
    \caption{Reduced free energy $G \ell^2/(\eta L)$ of the soliton branches as function of the magnetic flux $\Phi$. We set the spacelike period $\eta=0.5$. The green line is a new stable soliton branch.}
    \label{fig:alphaverysmall}
\end{figure}

To study phase transitions between the solutions, we need to match the asymptotic geometries, implying
\begin{eqnarray}
    \sqrt{g_{\tau_b\tau_b}(\rho \rightarrow \infty)} \beta_b &=& \sqrt{g_{\tau_s\tau_s}(\rho \rightarrow \infty)} \beta_s \\
     \sqrt{g_{\varphi_s\varphi_s}(\rho \rightarrow \infty)} \eta_s &=& \sqrt{g_{\varphi_b\varphi_b}(\rho \rightarrow \infty)} \eta_b
\end{eqnarray}
where $\rho$ is the cut-off in the radial direction. Since   for each configuration, the on-shell, regularized Euclidean actions provide a finite result, we can directly take the leading term in the expansion of the cutoff, leading to  $\beta_b= \beta_s = \beta$ and $\eta_s= \eta_b=\eta$. We also have to identify the electric potential of the black hole $\phi_b$ with that of the soliton, and the magnetic flux of the soliton $\Phi_s$ with that of the black hole.

In general phase transitions will occur between the planar black hole \eqref{planarbh} and the soliton \eqref{solbh} for a broad range of parameters and couplings. To illustrate this, we consider  the 3-horizon black hole,
for which  $b_{i\geq 5}=0$ 
implying 
\begin{equation}
\alpha_i= \frac{2^{i-1} (3i-3)!}{i! (2i-1)!} \alpha_2^{i-1}\quad 
\textrm{for $i \geq 3$}
\label{losalphas}
\end{equation}
In this case   the theory is characterized by a single NLED coupling $\alpha_2$, with the remaining infinite list of couplings   given by \eqref{losalphas}. For spherical black holes this choice of couplings also leads to very simple exact solutions \cite{Gao:2021kvr}, keeping analytic control over most of the thermal properties. Once the solution is constructed, one could apply the methods of \cite{Ayon-Beato:1999qin,Ayon-Beato:2004ywd} to obtain an NLED that accommodates such solutions.

Let us first analyze the many soliton branches that may appear for different values of the magnetic flux. The upper panel of figure \ref{fig:alphaverysmall} reproduces the result of \cite{Anabalon:2022ksf} in Maxwell electrodynamics for $\alpha_2=0$; the lower panel  shows that for small values of   $\alpha_2$, a new locally stable soliton branch emerges for a range of   magnetic flux. The dotted segments denote configurations for which the generalized specific heat is negative (the free energy is not a convex function of the magnetic flux).

\begin{figure}[H]
     \centering
     \begin{subfigure}[b]{0.3\textwidth}
         \centering
         \includegraphics[width=\textwidth]{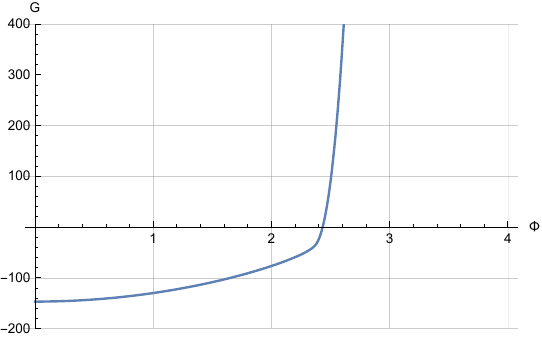}
         \caption{$\alpha_2=0.6$}
     \end{subfigure} 
     \qquad
     \begin{subfigure}[b]{0.3\textwidth}
         \centering
         \includegraphics[width=\textwidth]{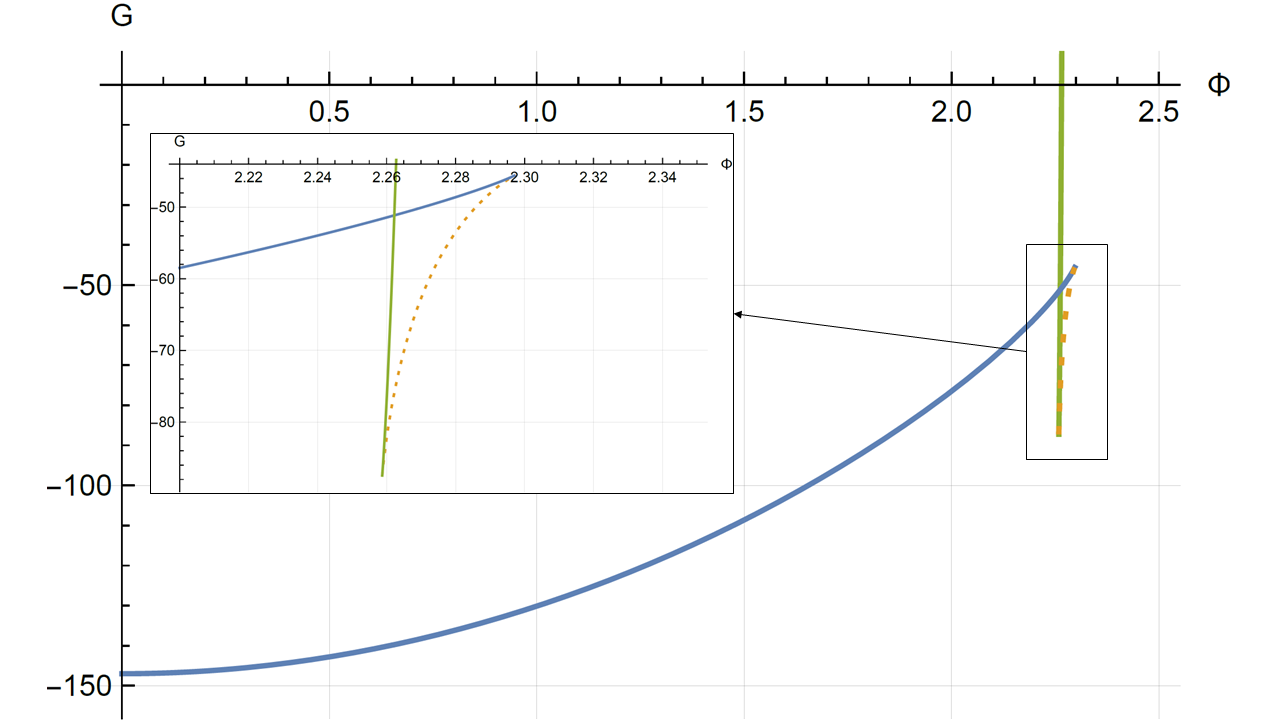}
         \caption{$\alpha_2=0.624$}
     \end{subfigure} \\
     
    \caption{Reduced free energy $G \ell^2/(\eta L)$ of the 
   small (blue) and large (green)
    soliton branches 
    as function of the magnetic flux $\Phi$ for larger values of $\alpha_2$. We set the spacelike period $\eta=0.5$.}
    \label{fig:alphasmall}
\end{figure}

As $\alpha_2$ increases the swallow-tail shrinks and then disappears. A new structure  appears with negative free energy when $\alpha_2\sim 0.625$, as shown in figure~\ref{fig:alphasmall}. We clearly see from the inset that when $\alpha_2=0.624$  a phase transition between solitons occurs, since for a given value of the the magnetic flux, the dominant soliton phase switches from small to large. Again, dotted segments represent configurations with non-convex free energy (negative specific heat). 

\begin{figure}[H]
\centering
\includegraphics[width=0.4\textwidth]{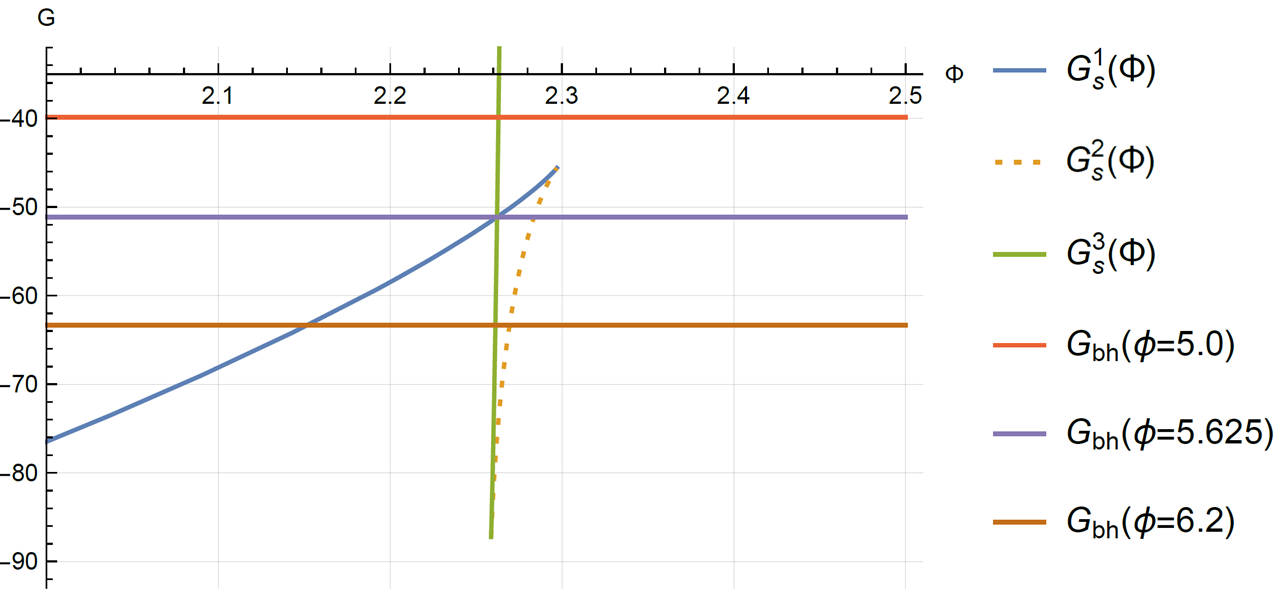}
\caption{$G \ell^2/(\eta L)$ for solitons and black hole solutions as function of the magnetic flux $\Phi$, for different, fixed values of the electric potential $\phi$. Here $\alpha_2=0.624$, $T=0.6$ and $\eta=0.5$.}
\label{fig:thept}
\end{figure}

\begin{figure}[H]
     \centering
     \begin{subfigure}[b]{0.4\textwidth}
         \centering
         \includegraphics[width=\textwidth]{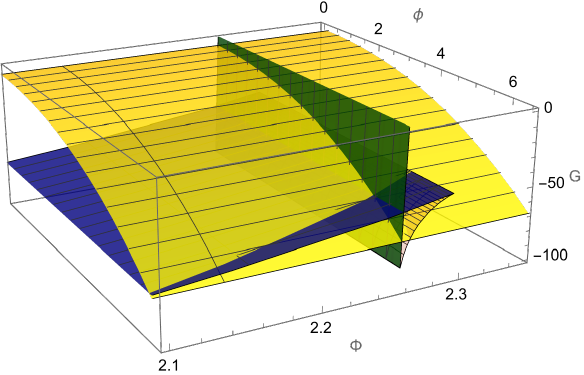}
     \end{subfigure}
       
    \caption{Reduced free energy $G \ell^2/(\eta L)$  for solitons (blue, orange and green surfaces) and black hole (yellow surface) as function of the magnetic flux $\Phi$ and the electric potential $\phi$. We set the nonlinear coupling $\alpha_2=0.624$, the temperature $T=0.6$ and the period of the spacelike cycle $\eta=0.5$.}
    \label{fig:surfaces}
\end{figure}

\textbf{The new triple Point and a reentrant phase transition}: Now that we have explored the soliton phase space, it is instructive to take the black hole phases into account. 
Figure \ref{fig:thept} shows that for large values of the electric potential $\phi$, there is a first order phase transition between the small soliton and the black hole, whereas for small values of $\phi$ this transition is between the black hole and the large soliton.
Significantly,
there is a precise value of $\phi$ and the magnetic flux $\Phi$, at which the free energy of the black hole (purple horizontal line), coincides with that of the two locally stable solitons (blue and green lines). At such a point in phase space the three configurations coexist, leading to a new type of triple point.
Notice also that there is a reentrant phase transition. For example at $\phi=6.2$, the small soliton (blue curve) dominates for   small $\Phi$; as $\Phi$ increases there is a first order transition to the black hole. Further increasing  $\Phi$, there is another first order transition to the large soliton (green curve).  Finally, above a second critical magnetic flux, the same black hole branch returns to being the dominant phase. The free energy of the black hole is insensitive to variations of the magnetic flux, while the free energy of the solitons is not sensitive to the electric potential. This can be see from plots of the surfaces $G(\phi,\Phi)$, shown in Figure \ref{fig:surfaces}. 
A more thorough exploration of the parameter space for this system will be provided elsewhere.

\section{Discussion}

We have shown the existence of a new type of triple point in General Relativity in AdS, supported by NLED. This point in phase space allows for the coexistence of two solitons and a black hole, at a given value of the temperature, magnetic flux, spacelike period and electric potential. Even though the NLED Lagrangian contains arbitrarily high powers of the Lorentz invariant $F^2$, both the electric and gravitational potential of the planar black hole are at most linear in the unique dimensionful coupling defining the NLED, as already observed for spherically symmetric black holes by Gao  \cite{Gao:2021kvr}. 

We emphasize that the behaviour we have observed  is generic for Einstein gravity coupled to any NLED and not simply the 3-horizon case considered here. In general multi-horizon black holes can be expected to exhibit multicritical soliton black hole phase behaviour, contingent on the number of horizons and the choice of parameters. This is because an increased number of horizons has an increased number of NLED couplings, yielding a  sufficient number of thermodynamic variables to support multiple phases \cite{Wu:2022bdk}.

Given the simple structure of the black holes in this theory, it would be natural to explore the existence of non-stationary configurations supported by this NLED, as recently done in \cite{Barrientos:2022bzm} for the case of MODMAX theory \cite{Bandos:2020jsw}, for accelerating black holes. We also expect that multicritical points,
similar to those observed for NLED black holes \cite{Tavakoli:2022kmo}, are also present for NLED solitons, and that perhaps extensions to more complicated solitons 
\cite{Andrews:2019hvq}
are possible. 
We expect to return to these topics in the near future, as well a to provide a more exhaustive exploration of the parameter space of the solitons and black holes.

\section{Acknowledgments} We thank Patrick Concha, Octavio Fierro and Evelyn Rodriguez for enlightening comments. This research has been supported in part by the Natural Sciences and Engineering Research Council of Canada, and
by FONDECYT grants 1221504, 1210635 and 1210500. This work was partially supported by ANID Fellowships 21191868
(C.Q.). The research of AA is supported in
part by a visiting researcher award of the FAPESP 2022/11765-7.

\bibliographystyle{fullsort.bst}
 
\bibliography{PT_NLE.bib}

\end{document}